\documentclass[preprint,showpacs,preprintnumbers,floatfix]{revtex4}

\usepackage{graphicx}
\usepackage{bm}

\def\lesssim{\ \raise.3ex\hbox{$<$}\kern-0.8em\lower.7ex\hbox{$\sim$}\ }
\def\gesim{\ \raise.3ex\hbox{$>$}\kern-0.8em\lower.7ex\hbox{$\sim$}\ }

\begin{document}

\title{Formation of magnetic impurities and pair-breaking effect in a superfluid Fermi gas}
\author{Y. Ohashi$^{1,2}$}
\affiliation{
$^1$Faculty of Science and Technology, Keio University, Hiyoshi, Yokohama, 223-8522, Japan\\
$^2$CREST(JST), 4-1-8 Honcho, Saitama 332-0012, Japan}
\date{\today}

\begin{abstract}
We theoretically investigate a possible idea to introduce magnetic impurities to a superfluid Fermi gas. In the presence of population imbalance ($N_\uparrow>N_\downarrow$, where $N_\sigma$ is the number of Fermi atoms with pseudospin $\sigma=\uparrow,\downarrow$), we show that {\it nonmagnetic} potential scatterers embedded in the system are {\it magnetized} in the sense that some of excess $\uparrow$-spin atoms are localized around them. They destroy the superfluid order parameter around them, as in the case of magnetic impurity effect discussed in the superconductivity literature. This pair-breaking effect naturally leads to localized excited states below the superfluid excitation gap. To confirm our idea in a simply manner, we treat an attractive Fermi Hubbard model within the mean-field theory at $T=0$. We self-consistently determine superfluid properties around a nonmagnetic impurity, such as the superfluid order parameter, local population imbalance, as well as single-particle density of states, in the presence of population imbalance. Since the competition between superconductivity and magnetism is one of the most fundamental problems in condensed matter physics, our results would be useful for the study of this important issue in cold Fermi gases.
\end{abstract}


\maketitle

\section{introduction}
Magnetic impurity effects have been extensively discussed in the field of metallic superconductivity. Within the Born approximation in terms of magnetic impurity scattering, Abrikosov and Gor'kov\cite{Abrikosov} (AG) showed that the superconducting phase is completely destroyed by magnetic impurities when the impurity concentration exceeds a critical value. Even below the critical impurity concentration, they showed that magnetic impurities remarkably affect superconducting properties, leading to the gapless superconductivity\cite{Abrikosov,Maki}, where the superconducting order parameter still exists but the BCS excitation gap is absent. We note that such a pair-breaking effect is absent in the case of nonmagnetic impurities, which is sometimes referred to as the Anderson's theorem\cite{Anderson}. 
\par
Shiba extended the AG theory to include multi-scattering processes beyond the Born approximation\cite{Shiba}. He clarified that magnetic impurities induce bound states below the BCS excitation gap $\Delta$. Using the numerical renormalization group theory, Shiba and co-workers further extended this theory to include the Kondo effect\cite{Shiba2,Shiba3}. They showed that the transition from the Kondo singlet to the spin doublet of magnetic impurity occurs at $T_{\rm K}/\Delta\simeq 0.3$, where $T_{\rm K}$ is the Kondo temperature. 
\par
The recently realized superfluid Fermi gases in $^{40}$K\cite{Regal} and $^6$Li\cite{Bartenstein,Zwierlein,Kinast} have the unique property that one can tune the strength of a pairing interaction by adjusting the threshold energy of a Feshbach resonance. Using this advantage, we can now study superfluid properties from the weak-coupling BCS (Bardeen-Cooper-Schrieffer) regime to the strong-coupling BEC (Bose-Einstein condensation) regime in a unified manner\cite{Chen,Eagles,Leggett,NSR,SadeMelo,Haussmann,Ohashi,Strinati,Haussmann2,Haussmann3,Hu}. In addition, since the cold Fermi gas system is much cleaner and simpler than metallic superconductors, the former is expected as a useful quantum simulator to study various phenomena observed in the latter complicated systems, without being disturbed by extrinsic effects. Indeed, the pseudogap phenomenon, which has been extensively discussed in high-$T_{\rm c}$ cuprates\cite{Review,Randeria,Singer,Janko,Rohe,Yanase}, has been recently observed in $^{40}$K Fermi gases\cite{Jin1,Jin2}. Since the latter system is dominated by strong-pairing fluctuations, we can now study how the pseudogap is induced by pairing fluctuations in a clear manner\cite{Tsuchiya,Watanabe,Hu2}. Thus, if one could study magnetic impurity effects in superfluid Fermi gases, it would be helpful for further understanding of the competition between superconductivity and magnetism, which is one of the most fundamental problems in condensed matter physics.
\par
\begin{figure}[t]
\includegraphics[width=8cm]{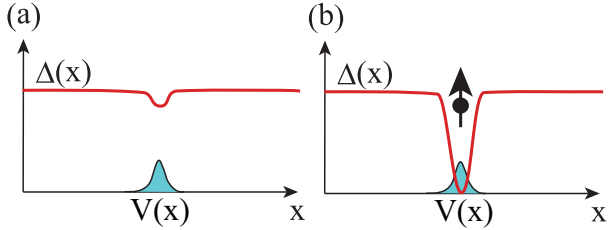}
\caption{(Color online) Schematic picture of our idea to realize magnetic impurities. $V(x)$ is a weak nonmagnetic impurity potential embedded in a superfluid Fermi gas. (a) In an unpolarized Fermi superfluid, although the nonmagnetic impurity does not destroy the superfluid state, the superfluid order parameter $\Delta(x)$ is slightly suppressed around the impurity, because of the decrease of particle density by the impurity potential $V(x)$. (b) When an $\uparrow$-spin atom (solid circle with $\uparrow$) is added to the system, this excess atom is localized around the impurity, so as to minimize the loss of condensation energy by the pair-breaking effect. As a result, the impurity becomes magnetic, and the superfluid order parameter is remarkably suppressed around it. 
\label{fig1}
}
\end{figure}

In this paper, we theoretically discuss an idea to realize magnetic impurities in a superfluid Fermi gas. So far, {\it nonmagnetic} impurities have been realized in cold atom gases by using another species of atoms\cite{Zipkes} and laser light\cite{Modugno}. One difficulty in realizing magnetic impurities in a superfluid Fermi gas is that spins ($\sigma=\uparrow,\downarrow$) in this system are actually, not real spins, but `pseudo'-spins describing two atomic hyperfine states. Thus, magnetic impurities must be also `pseudo'-magnetic objects acting on these atomic hyperfine states. In this paper, to realize such pseudo-magnetic scatterers, we use nonmagnetic impurities as seeds. Although the nonmagnetic impurities do not destroy superfluid state (Anderson's theorem), the superfluid order parameter may slightly decrease around the impurities, because of the slight suppression of particle density, as schematically shown in Fig.\ref{fig1}(a). In this case, when $\uparrow$-spin atoms are added to the system, since they cannot form Cooper-pairs, they would behave like polarized magnetic impurities, and destroy the superfluid order parameter around them. Thus, in order to minimize the loss of condensation energy by this pair-breaking effect, as shown in Fig.\ref{fig1}(b), excess $\uparrow$-spin atoms are expected to be localized around the non-magnetic impurities, because the superfluid order parameter around the impurities has been already slightly suppressed before the $\uparrow$-spin doping. The resulting nonmagnetic impurities accompanied by localized $\uparrow$-spin atoms may be viewed as magnetic impurities.

\par
To confirm this scenario, in this paper, we consider a two-dimensional attractive Fermi Hubbard model with a nonmagnetic impurity. Although this simple lattice model is different from the real three-dimensional continuum Fermi gas system, we emphasize that the presence of lattice, as well as the low-dimensionality, are not essential for the present problem. Within the framework of mean-field theory at $T=0$, we calculate the spatial variation of the superfluid order parameter, as well as local population imbalance, around the impurity. In a polarized Fermi superfluid, we confirm that the nonmagnetic impurity is really magnetized. We also show that the superfluid order parameter is damaged by this magnetized impurity, leading to bound states below the BCS excitation gap.
\par
We note that a different idea to realize magnetic impurities has been recently proposed in Ref.\cite{Demler}, where an impurity with $a_\uparrow\ne a_\downarrow$ is used (where $a_\sigma$ is a scattering length between an impurity and an atom with pseudospin $\sigma$). We also note that, under the assumption of the presence of a magnetic impurity, properties of bound states have been examined in Ref.\cite{Pu}.
\par
The outline of this paper is as follows: In Sec. II, we explain our formulation to calculate superfluid properties around a nonmagnetic impurity. In Sec. III, we numerically confirm our idea about the formation of magnetic impurities in a polarized Fermi gas. We discuss bound states induced by the magnetized impurity in Sec. IV. In Sec. V, we briefly examine effects of a trap potential, as well as finite temperatures. Throughout this paper, we set $\hbar=k_{\rm B}=1$, and take the lattice constant unity.
\par
\section{Formulation}
\par
We consider a two-component Fermi gas, described by pseudospin $\sigma=\uparrow,\downarrow$. Assuming a two-dimensional square lattice, we put a nonmagnetic impurity at the center of the system. For simplicity, we first ignore effects of a trap potential, which will be separately examined in Sec.V. The model Hamiltonian is given by
\begin{equation}
{\hat H}=-t\sum_{\langle i,j\rangle,\sigma}
\left[
{\hat c}_{i,\sigma}^\dagger {\hat c}_{j,\sigma}+h.c
\right]
-U\sum_i {\hat n}_{i,\uparrow}{\hat n}_{i,\downarrow}
+\sum_{i,\sigma}
\left[V(i)-\mu_\sigma\right] {\hat n}_{i,\sigma},
\label{eq.1}
\end{equation}
where ${\hat c}_{i,\sigma}^\dagger$ is the creation operator of a Fermi atom with pseudospin $\sigma$ at the $i$-th site. ${\hat n}_{i,\sigma}={\hat c}_{i,\sigma}^\dagger {\hat c}_{i,\sigma}$ is the number operator at the $i$-th site. $-t$ describes atomic hopping between nearest-neighbor sites, and the summation $\langle i,j\rangle$ is taken over nearest-neighbor pairs. $-U$ is an on-site pairing interaction between $\uparrow$-spin atom and $\downarrow$-spin atom. Since we consider a polarized Fermi superfluid, the chemical potential $\mu_\sigma$ depends on $\sigma=\uparrow,\downarrow$. The nonmagnetic impurity potential $V(i)$ is assumed to have the form
\begin{equation}
V(i)
=
V_0 {\Gamma^2 \over (R_x-R_{x0})^2+(R_y-R_{y0})^2+\Gamma^2},
\label{eq.1b}
\end{equation}
where $(R_x,R_y)$ is the spatial position of the $i$-th lattice site. $(R_{x0},R_{y0})$ is the center of the impurity potential.
\par
We treat Eq. (\ref{eq.1}) within the mean-field theory at $T=0$. Introducing the superfluid order parameter $\Delta(i)=U\langle {\hat c}_{i,\downarrow}{\hat c}_{i,\uparrow}\rangle$ (which is taken to be real in this paper), as well as the particle density $n_\sigma(i)=\langle {\hat n}_{i,\sigma}\rangle$, we obtain the mean-field Hamiltonian ${\hat H}_{\rm MF}$ as
\begin{eqnarray}
{\hat H}_{\rm MF}
&=&-t\sum_{\langle i,j\rangle,\sigma}
\left[{\hat c}_{i,\sigma}^\dagger {\hat c}_{j,\sigma}+h.c.\right]
-\sum_i
\Delta(i)\left[{\hat c}_{i,\uparrow}^\dagger {\hat c}_{i,\downarrow}^\dagger+h.c.
\right]
\nonumber
\\
&+&
\sum_i[{\tilde V}(i)-\mu]{\hat n}_{i,\sigma}
-h\sum_i{\hat s}_{z,i}
+
{U \over 2}\sum_is_z(i){\hat s}_{z,i}.
\label{eq.2}
\end{eqnarray}
In Eq. (\ref{eq.2}), we have ignored unimportant constant terms. (We also ignore them in the following discussions.) ${\hat s}_{z,i}={\hat n}_{i,\uparrow}-n_{i,\downarrow}$ is the spin density operator, and
\begin{equation}
s_z(i)=n_\uparrow(i)-n_\downarrow(i)
\label{eq.16}
\end{equation}
represents the local magnetization. When this quantity becomes finite around the impurity, the last term in Eq. (\ref{eq.2}) works as an Ising-type magnetic impurity scatterer. The effective impurity potential ${\tilde V}(i)=V(i)+V_{\rm Hartree}(i)$ involves the inhomogeneous Hartree term, $V_{\rm Hartree}(i)=-(U/2)[n_\uparrow(i)-n_\downarrow(i)-n_0]$, where $n_0$ is the particle density far away from the impurity. $\mu=\mu_{\rm av}+(U/2)n_0$ is an effective chemical potential, where $\mu_{\rm av}=[\mu_\uparrow+\mu_\downarrow]/2$. The difference of the spin-dependent chemical potentials
\begin{equation}
h={1 \over 2}[\mu_\uparrow-\mu_\downarrow]
\label{eq.h}
\end{equation}
works as an effective magnetic field in Eq. (\ref{eq.2}). 
\par
Since Eq. (\ref{eq.2}) has a bi-linear form, we can write it in the form ${\hat H}_{\rm MF}={\hat \Phi}^\dagger{\tilde H}_{\rm MF}{\hat \Phi}$. Here, ${\hat \Phi}^\dagger=({\hat c}_{1,\uparrow}^\dagger,{\hat c}_{2,\uparrow}^\dagger,\cdots, {\hat c}_{M,\uparrow}^\dagger,{\hat c}_{1,\downarrow},{\hat c}_{2,\downarrow},\cdots,{\hat c}_{M,\downarrow})$, where $M$ is the total number of lattice sites. The Hamiltonian matrix ${\tilde H}_{\rm MF}$ is chosen so as to reproduce Eq. (\ref{eq.2}). As usual, Eq. (\ref{eq.2}) can be diagonalized by the Bogoliubov transformation,
\begin{eqnarray}
{\hat \Phi}=
{\hat W}
\left(
\begin{array}{c}
{\hat \gamma}_{1}\\
{\hat \gamma}_{2}\\
:\\
{\hat \gamma}_{M}\\
{\hat \gamma}_{M+1}\\
{\hat \gamma}_{M+2}\\
:\\
{\hat \gamma}_{2M}\\
\end{array}
\right).
\label{eq.4}
\end{eqnarray}
The $2M\times 2M$ orthogonal matrix ${\hat W}$ is determined so as to diagonalize ${\tilde H}_{\rm MF}$. The diagonalized mean-field Hamiltonian has the form
\begin{equation}
{\hat H}_{\rm MF}=\sum_{j=1}^{2M}E_j{\hat \gamma}_j^\dagger{\hat \gamma}_j,
\label{eq.5}
\end{equation}
where $E_j$ is the $j$-th eigen energy. 
\par
The superfluid order parameter $\Delta(i)$, as well as the particle density $\langle n_{i,\sigma}\rangle$ are given by, respectively,
\begin{equation}
\Delta(i)=U\sum_{j=1}^{2M}W_{i,j}W_{M+i,j}\Theta(-E_j),
\label{eq.6}
\end{equation}
\begin{equation}
n_\uparrow(i)=
\sum_{j=1}^{2M}W_{i,j}^2\Theta(-E_j),
\label{eq.7}
\end{equation}
\begin{equation}
n_\downarrow(i)=
\sum_{j=1}^{2M}W_{M+i,j}^2\Theta(E_j),
\label{eq.8}
\end{equation}
where $\Theta(x)$ is the step function. The chemical potential $\mu_\sigma$ is determined from the equation for the total number of $\sigma$-spin atoms, 
\begin{equation}
N_\sigma=\sum_i n_\sigma(i).
\label{eq.8b}
\end{equation} 
\par

\begin{figure}[t]
\includegraphics[width=8cm]{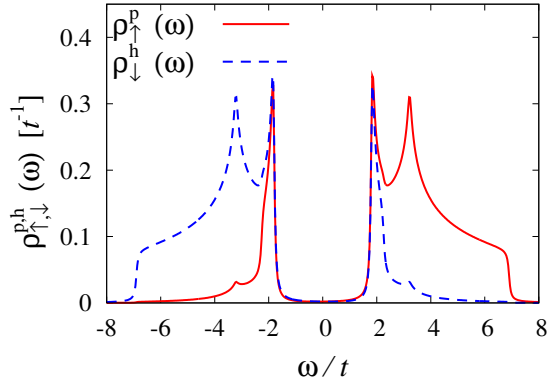}
\caption{(Color online) Superfluid density of states in an unpolarized uniform BCS superfluid. The solid line and dashed line show the $\uparrow$-spin component $\rho_\uparrow^{\rm p}(\omega)$ and $\downarrow$-spin component $\rho_\downarrow^{\rm h}(\omega)$, respectively. In this figure, we have use the same values of the superfluid order parameter $\Delta$ and chemical potential $\mu$ as $\Delta(1,1)$ and $\mu_\sigma$ in Fig.\ref{fig5}, respectively.
\label{fig2}
}
\end{figure}
\par
To examine single-particle properties around the impurity, we consider the local density of states (LDOS). To calculate this, we introduce the single-particle thermal Green's function for $\uparrow$-spin atoms,
\begin{equation}
G^{\rm p}_{\uparrow\uparrow}(i,i,i\omega_n)=
-\int_0^\beta d\tau e^{i\omega\tau}\langle
T_\tau
\{
{\hat c}_{i,\uparrow}(\tau){\hat c}_{i,\uparrow}^\dagger(0)
\}
\rangle,
\label{eq.9}
\end{equation}
as well as the Green's function for $\downarrow$-spin holes,
\begin{equation}
G^{\rm h}_{\downarrow\downarrow}(i,i,i\omega_n)=
-\int_0^\beta d\tau e^{i\omega\tau}\langle
T_\tau
\{
{\hat c}^\dagger_{i,\downarrow}(\tau){\hat c}_{i,\downarrow}(0)
\}
\rangle.
\label{eq.10}
\end{equation}
Here, $\omega_n$ is the fermion Matsubara frequency, and $\beta=1/T$ is the inverse temperature. In the mean-field theory, Eq. (\ref{eq.9}) is given by
\begin{eqnarray}
G^{\rm p}_{\uparrow\uparrow}(i,i,i\omega_n)
&=&
-\sum_{j,j'=1}^{2M} W_{i,j}W_{i,j'}
\int_0^\beta d\tau e^{i\omega\tau}\langle
T_\tau
\{
{\hat \gamma}_{j}(\tau){\hat \gamma}_{j'}^\dagger(0)
\}
\rangle
\nonumber
\\
&=&
\sum_{j=1}^{2M} {W_{i,j}^2 \over i\omega_n-E_j}.
\label{eq.11}
\end{eqnarray}
Executing the analytic continuation in Eq. (\ref{eq.11}), we obtain LDOS for the $\uparrow$-spin component as
\begin{eqnarray}
\rho^p_{i,\uparrow}(\omega)
&=&
-{1 \over \pi}{\rm Im}[G^{\rm p}_{\uparrow\uparrow}(i,i,i\omega_n\to\omega+i\delta)]
\nonumber
\\
&=&
\sum_{j=1}^{2M}W_{i,j}^2\delta(\omega-E_j).
\label{eq.12}
\end{eqnarray}
In the same manner, LDOS for the $\downarrow$-spin component is obtained from the analytic continuation of Eq. (\ref{eq.10}), as
\begin{eqnarray}
\rho^h_{i,\downarrow}(\omega)
&=&
-{1 \over \pi}{\rm Im}[G^{\rm h}_{\downarrow\downarrow}(i,i,i\omega_n\to\omega+i\delta)]
\nonumber
\\
&=&
\sum_{j=1}^{2M}W_{M+i,j}^2\delta(\omega-E_j).
\label{eq.13}
\end{eqnarray}
\par
In the absence of nonmagnetic impurity and population imbalance, Eq. (\ref{eq.12}) reduces to the ordinary superfluid density of states in the uniform BCS state,
\begin{equation}
\rho^{\rm p}_\uparrow(\omega)=
\sum_{\bf p}u_{\bf p}^2
\delta(\omega-E_{\bf p})
+
\sum_{\bf p}v_{\bf p}^2
\delta(\omega+E_{\bf p}),
\label{eq.18a}
\end{equation}
and Eq. (\ref{eq.13}) also reduces to the BCS expression,
\begin{equation}
\rho^{\rm h}_\downarrow(\omega)=
\sum_{\bf p}u_{\bf p}^2
\delta(\omega+E_{\bf p})
+
\sum_{\bf p}v_{\bf p}^2
\delta(\omega-E_{\bf p}),
\label{eq.18b}
\end{equation}
where 
\begin{eqnarray}
u_{\bf p}^2={1 \over 2}
\Bigl[
1+{\xi_{\bf p} \over E_{\bf p}}
\Bigr],
\label{eq.18c}
\\
v_{\bf p}^2={1 \over 2}
\Bigl[
1-{\xi_{\bf p} \over E_{\bf p}}
\Bigr].
\label{eq.18d}
\end{eqnarray}
In Eqs.(\ref{eq.18c}) and (\ref{eq.18d}), $E_{\bf p}=\sqrt{\xi_{\bf p}^2+\Delta^2}$ is the Bogoliubov excitation spectrum, where $\Delta$ is the uniform superfluid order parameter. $\xi_{\bf p}=\varepsilon_{\bf p}-\mu$ is the kinetic energy $\varepsilon_{\bf p}=-2t[\cos p_x+\cos p_y]$ of the tight-binding model, measured from the chemical potential $\mu$. As shown in Fig.\ref{fig2}, $\rho_\uparrow^{\rm p}(\omega)$ and $\rho_\downarrow^{\rm h}(\omega)$ have the BCS coherence peaks at $\omega\simeq \pm1.8 t$, as well as the van Hove singularities associated with the two-dimensional square lattice model at $\omega\simeq\pm 3.2t$. From Eqs. (\ref{eq.18a}) and (\ref{eq.18b}), one finds that $\rho_{i,\uparrow}^{\rm p}(\omega)$ describes $\uparrow$-spin particle (hole) excitations when $\omega>0$ ($\omega<0$). $\rho_{i,\downarrow}^{\rm h}(\omega)$ describes $\downarrow$-spin hole (particle) excitations when $\omega>0$ ($\omega<0$).
\par
We note that these physical meanings of $\rho_{i,\uparrow}^{\rm p}(\omega)$ and $\rho_{i,\downarrow}^{\rm h}(\omega)$ are slightly altered in the polarized case ($\mu_\uparrow\ne\mu_\downarrow$). In this case, within the neglect of the phase separation phenomenon, we again obtain Eqs. (\ref{eq.18a}) and (\ref{eq.18b}), where the energy $\omega$ is replaced by $\omega+h$ (where $h$ is given in Eq. (\ref{eq.h})). Thus, in the polarized case, we need to replace the energy $\omega$ by $\omega+h$ in the above discussions about the physical meanings of $\rho_{i,\uparrow}^{\rm p}(\omega)$ and $\rho_{i,\downarrow}^{\rm h}(\omega)$.
\par
We also note that Eqs. (\ref{eq.12}) and (\ref{eq.13}) indicate that $W_{i,j}^2$ and $W_{M+i,j}^2$, respectively, describe the probability densities of $\uparrow$-spin and $\downarrow$-spin components of the $j$-th eigen state at the $i$-th site. When we write the wavefunction of the $j$-th eigenstate in the form $\Psi_j(i)=(\Psi_j^{\rm p}(i),\Psi_j^{\rm h}(i))$, where each component is related to $W_{i,j}^2$ and $W_{M+i,j}^2$ as
\begin{eqnarray}
\left\{
\begin{array}{l}
|\Psi_j^p(i)|^2=W_{i,j},\\
|\Psi_j^h(i)|^2=W_{M+i,j},
\end{array}
\right.
\label{eq.14} 
\end{eqnarray}
this wavefunction is normalized as (Note that ${\hat W}$ is an orthogonal matrix.)
\begin{equation}
\sum_{i=1}^M
\left[
|\Psi_j^p(i)|^2+|\Psi_j^h(i)|^2
\right]
=\sum_{i=1}^{2M}W_{i,j}^2=1.
\label{eq.15}
\end{equation}
From the physical meanings of $\rho_{i,\uparrow}^{\rm p}(\omega)$ and $\rho_{i,\downarrow}^{\rm h}(\omega)$, one may interpret $\Psi_j^{\rm p}(\omega)$ and $\Psi_j^h(i)$ as the $\uparrow$-spin particle and $\downarrow$-spin hole components, respectively, when the eigen energy $\omega$ is larger than $-h$. When $\omega<-h$, $\Psi_j^{\rm p}(\omega)$ and $\Psi_j^h(i)$ have the physical meanings of $\uparrow$-spin hole and $\downarrow$-spin particle components, respectively. We will use these interpretations in Sec. IV, when we examine physical properties of bound states. 
\par

\begin{figure}[t]
\includegraphics[width=8cm]{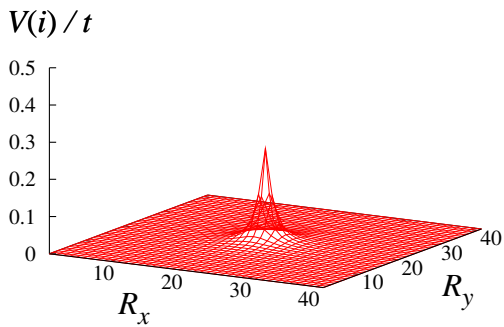}
\caption{(Color online) Model nonmagnetic impurity potential $V(i)$ used in this paper. We take $(R_{0x},R_{0y})=(21,21)$, $V_0/t=0.25$, and $\Gamma=1$.
\label{fig3}
}
\end{figure}
\par
Before ending this section, we summarize details of our numerical calculations. We consider a $41\times 41$ square lattice, and impose the periodic boundary condition. For the magnitude of on-site pairing interaction $U$, we take $U/t=6$. To avoid band effects originating from the nested Fermi surface near the half-filling, we consider the low density region, by setting $N_\uparrow=N_\downarrow=300$ in the absence of population imbalance (which gives the the particle density per lattice site $[N_\uparrow+N_\downarrow]/M=0.357\ll 1$). Starting from this unpolarized case, we gradually increase the number of $\uparrow$-spin atoms to explore the possibility of localization of excess atoms around the nonmagnetic impurity at $(R_{x0},R_{y0})=(21,21)$. For the impurity potential, we take $V_0/t=0.25$, and $\Gamma=1$ in the unit of lattice constant. The resulting impurity potential $V(i=(R_x,R_y))$ is shown in Fig.\ref{fig3}. 
\par
Under these conditions, we numerically diagonalize ${\tilde H}_{\rm MF}$ to obtain ${\hat W}$ in Eq. (\ref{eq.4}), for a given set of ($\Delta(i)$, $n_\uparrow(i)$, $n_\downarrow(i)$, $\mu_\uparrow$, $\mu_\downarrow$). We update them by using Eqs. (\ref{eq.6})-(\ref{eq.8b}), until the self-consistency is achieved. We then calculate superfluid properties around the impurity, such as LDOS's in Eqs. (\ref{eq.12}) and (\ref{eq.13}), as well as wavefunctions of low-lying excited states in Eq. (\ref{eq.14}).

\begin{figure}[t]
\includegraphics[width=8cm]{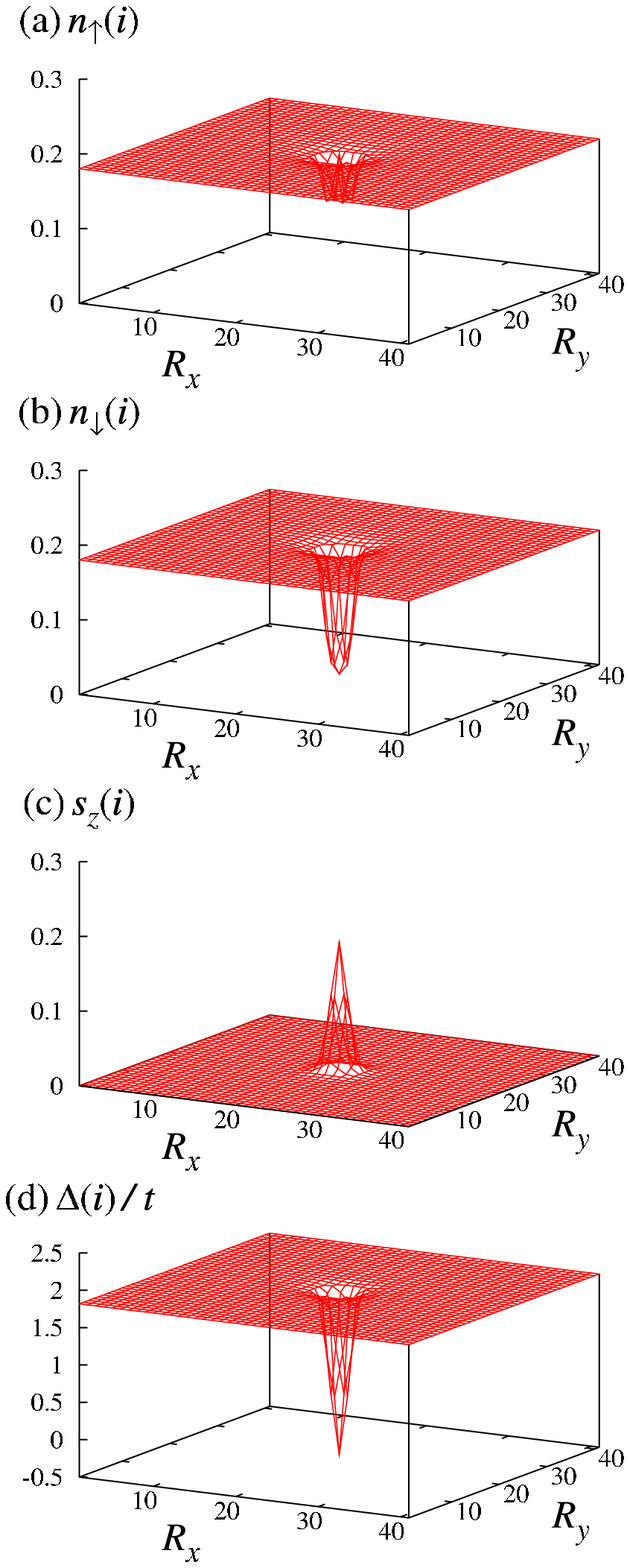}
\caption{(Color online) Calculated superfluid state around the nonmagnetic impurity at $(R_{x},R_{y})=(21,21)$. We take the population imbalance as $N_\uparrow=301>N_\downarrow=300$, which gives the spin-dependent chemical potentials, $\mu_\uparrow=-2.26t$ and $\mu_\downarrow=-5.20t$. Panels (a) and (b) show the particle densities of $\uparrow$-spin atoms and $\downarrow$-spin atoms, respectively. (c) Magnetization $s_z(i)=n_\uparrow(i)-n_\downarrow(i)$. (d) Superfluid order parameter $\Delta(i)$. 
\label{fig4}
}
\end{figure}

\begin{figure}[t]
\includegraphics[width=8cm]{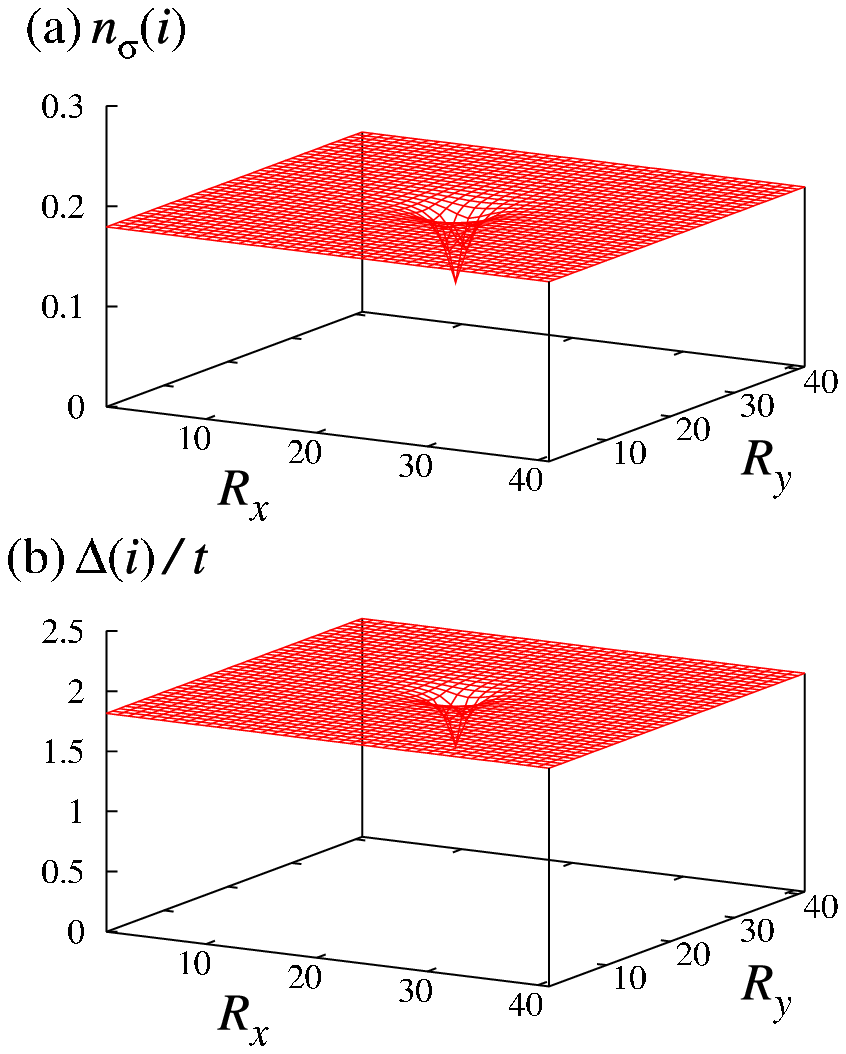}
\caption{(Color online) Superfluid state around the nonmagnetic impurity at $(R_{x},R_{y})=(21,21)$, in the unpolarized case ($N_\uparrow=N_\downarrow=300$). In this case, we obtain $\mu_\uparrow=\mu_\downarrow=-3.73t$. (a) $n_\sigma(i)$ ($\sigma=\uparrow,\downarrow$). (b) $\Delta(i)$. We note that $s_z(i)=0$ in the unpolarized case. 
\label{fig5}
}
\end{figure}

\section{Formation of magnetic impurity and pair-breaking effect}
\par 
Figure \ref{fig4} shows the superfluid properties around the nonmagnetic impurity in the polarized case ($N_\uparrow=301>N_\downarrow=300$). For comparison, we also show the results in the unpolarized case in Fig.\ref{fig5} ($N_\uparrow=N_\downarrow=300$). As expected, Figs.\ref{fig4}(a)-(c) show that the nonmagnetic impurity is {\it magnetized} in the sense that an excess $\uparrow$-spin atom is localized around it. Then, the last term in Eq. (\ref{eq.2}) works as a magnetic impurity potential. Indeed, Fig.\ref{fig4}(d) shows that the superfluid order parameter $\Delta(i)$ is damaged around the impurity, as in the case of ordinary magnetic impurity effect in metallic superconductivity\cite{Abrikosov,Shiba}. 
\par
In the unpolarized case shown in Fig.\ref{fig5}, such a strong depairing effect is not obtained. The slight decrease of $\Delta(i)$ around the impurity is only seen in Fig.\ref{fig5}(b), reflecting the slight suppression of the particle density $n_\sigma(i)$ shown in Fig.\ref{fig5}(a). This confirms that the remarkable suppression of $\Delta(i)$ in Fig.\ref{fig4}(d) is attributed to, not the original impurity potential $V(i)$, but the localized $\uparrow$-spin atom. The role of the weak nonmagnetic impurity potential $V(i)$ is only to slightly decrease the superfluid order parameter around it, so as to capture the excess atom.
\par
\begin{figure}[t]
\includegraphics[width=8cm]{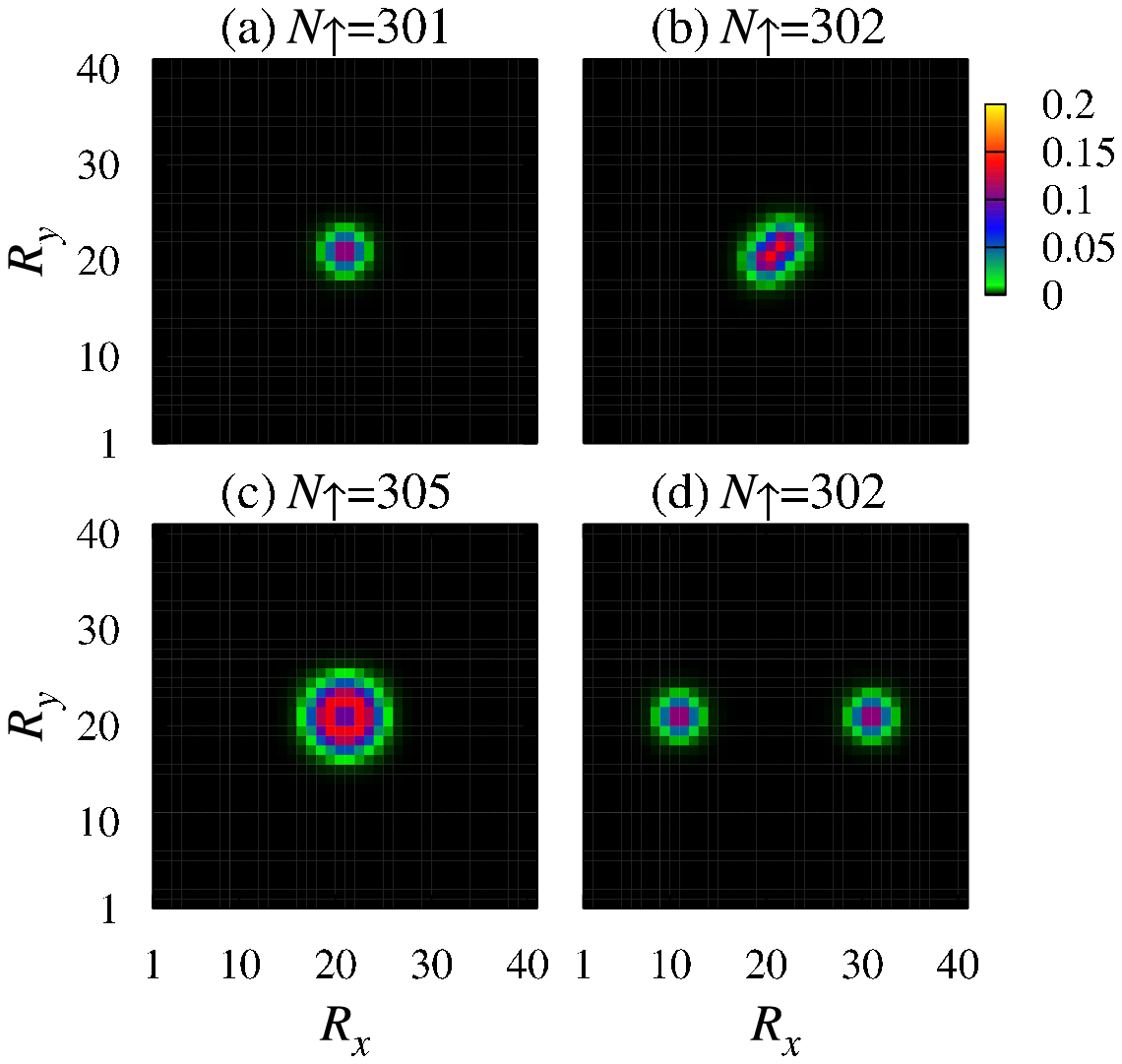}
\caption{(Color online) Intensity of the local magnetization $s_z(i)$ around the impurity put at the center of the system. We take $N_\downarrow=300$. In panel (d), two impurities are put at $(R_{x},R_{y})=(11,21)$ and $(31,21)$. The spatial structure of each impurity potential is the same as Eq. (\ref{eq.1b}).
\label{fig6}
}
\end{figure}

When we increase the number of excess $\uparrow$-spin atoms, they cluster around the impurity, so that local magnetization $s_z(i)$ around the impurity increases, as shown in Figs.\ref{fig6}(a)-(c)\cite{note}. In addition, in the case of two excess atoms shown in Fig.\ref{fig6}(b), when we put one more impurity in the system, we obtain two magnetic impurities, as shown in Fig.\ref{fig6}(d), where  each impurity is accompanied by one excess $\uparrow$-spin atom. Thus, using these, one can tune the magnitude of magnetization of each impurity, as well as the concentration of magnetic impurities. 
\par

\begin{figure}
\includegraphics[width=8cm]{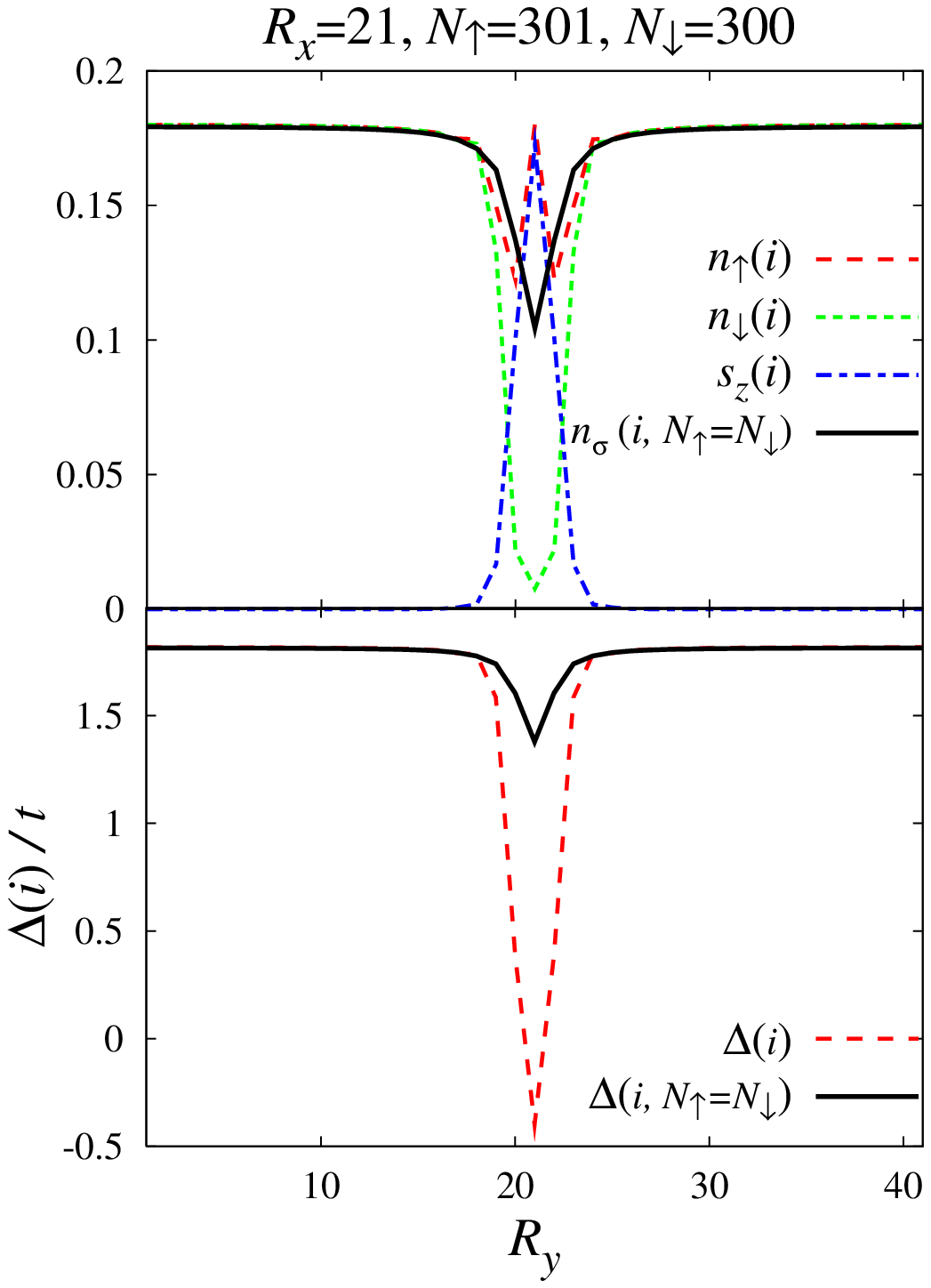}
\caption{(Color online) (a) Particle density $n_\sigma(i)$ ($\sigma=\uparrow,\downarrow$) at $R_x=21$, as a function of $R_y$. We take $N_\uparrow=301$ and $N_\downarrow=300$. In this figure, $n_\sigma(i,N_\uparrow=N_\downarrow)$ is the particle density in the unpolarized case ($N_\uparrow=N_\downarrow=300$). (b) Spatial variation of the superfluid order parameter $\Delta(i)$ at $R_x=21$. $\Delta(i,N_\uparrow=N_\downarrow)$ is the result in the absence of population imbalance ($N_\uparrow=N_\downarrow=300$). 
\label{fig7}
}
\end{figure}
\par
Here, we comment on the similarity between the magnetization of nonmagnetic impurity in Fig.\ref{fig4} and the phase separation observed in polarized Fermi gases\cite{MIT3,RICE}. In the latter phenomenon, the system spatially separate into the unpolarized superfluid region and polarized normal region, and the minority $\downarrow$-spin component is almost excluded from the latter region\cite{MIT3,RICE,Silva,Chevy,Pieri,Yi,Silva2,Haque}. Such exclusion of $\downarrow$-spin component from the polarized region can be also seen in the present magnetic impurity problem. Indeed, Fig.\ref{fig7}(a) shows that, while the density profile $n_\uparrow(i)$ of $\uparrow$-spin atoms is not so different from that in the unpolarized case (except for the enhancement at the impurity site ($R_y=21$) due to the localization of the excess $\uparrow$-spin atom), $\downarrow$-spin atoms are clearly pushed out from the region around the impurity ($R_y\simeq 21$).
\par
Another similarity between the present magnetic impurity formation and the phase separation in polarized Fermi gases is the {\it negative} superfluid order parameter ($\Delta(i)<0$) around the impurity seen in Fig.\ref{fig7}(b). In a polarized Fermi superfluid, the possibility of Fulde-Ferrell-Larkin-Ovchinnikov (FFLO) state (which is characterized by the spatial oscillation of the superfluid order parameter) has been proposed near the boundary between the superfluid region and polarized normal region\cite{Machida}. The oscillation of the FFLO order parameter is the direct consequence of the formation of Cooper pairs with a finite momentum, originating from the mismatch of Fermi surfaces between the $\uparrow$-spin component and $\downarrow$-spin component in the polarized region. Although the polarized region is spatially small in the case of Fig.\ref{fig7}, one can still expect an FFLO-like pairing in the magnetized region, which leads to the sign change of the superfluid order parameter seen in Fig.\ref{fig7}(b).
\par

\begin{figure}[t]
\includegraphics[width=8cm]{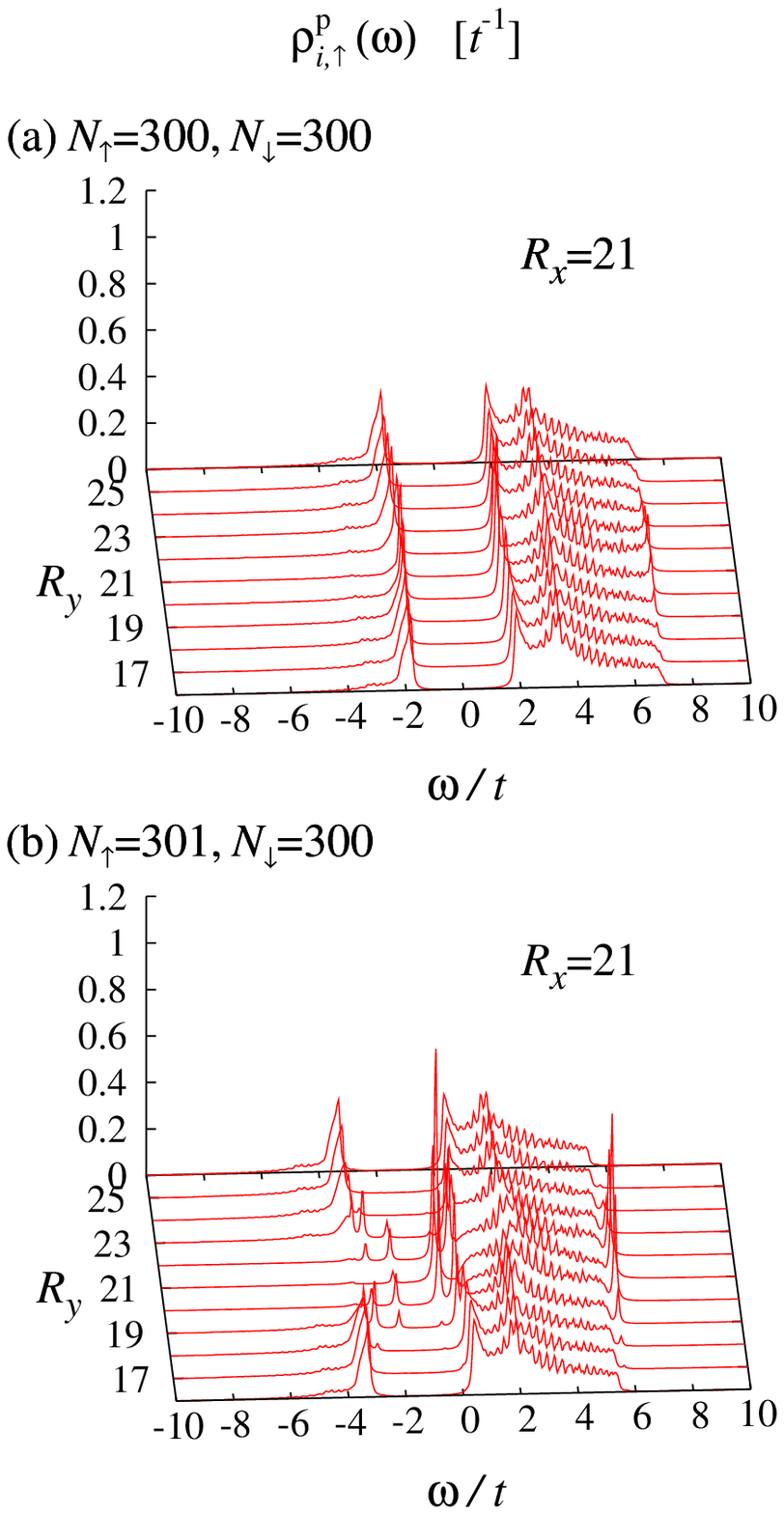}
\caption{(Color online) Superfluid local density of states (LDOS) $\rho^{\rm p}_{i,\uparrow}(\omega)$ around the impurity. In this figure, results at $R_y=21$ show LDOS at the impurity site. (a) Unpolarized case ($N_\uparrow=N_\downarrow=300$). (b) Polarized case ($N_\uparrow=301>N_\downarrow=300$), where one excess $\uparrow$-spin atom is localized around the impurity. In calculating LDOS, we have added the small imaginary part $\delta=0.05t$ to eigen energies, in order to smear out unphysical fine structures around the coherence peaks, coming from discrete energy levels in the finite system. We briefly note that the influence of discrete levels still remains as sharp peaks around the van Hove singularity. We also use this prescription in Figs.\ref{fig9} and \ref{fig10}. 
\label{fig8}
}
\end{figure}

\begin{figure}[t]
\includegraphics[width=8cm]{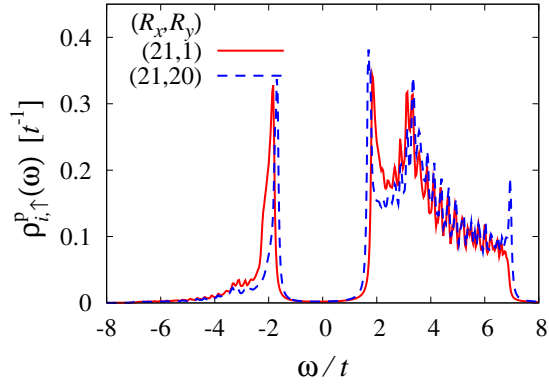}
\caption{
(Color online) Local density of states $\rho^{\rm p}_{i,\uparrow}(\omega)$ in the unpolarized case ($N_\uparrow=N_\downarrow=300$). Apart from the fine peak structure seen around the van Hove singularity (which is due to the discrete levels in the finite system), LDOS far away from the impurity (solid line) agrees with the uniform BCS density of states $\rho^{\rm p}_\uparrow(\omega)$ in Fig.\ref{fig2}. 
\label{fig9}
}
\end{figure}

\begin{figure}[t]
\includegraphics[width=8cm]{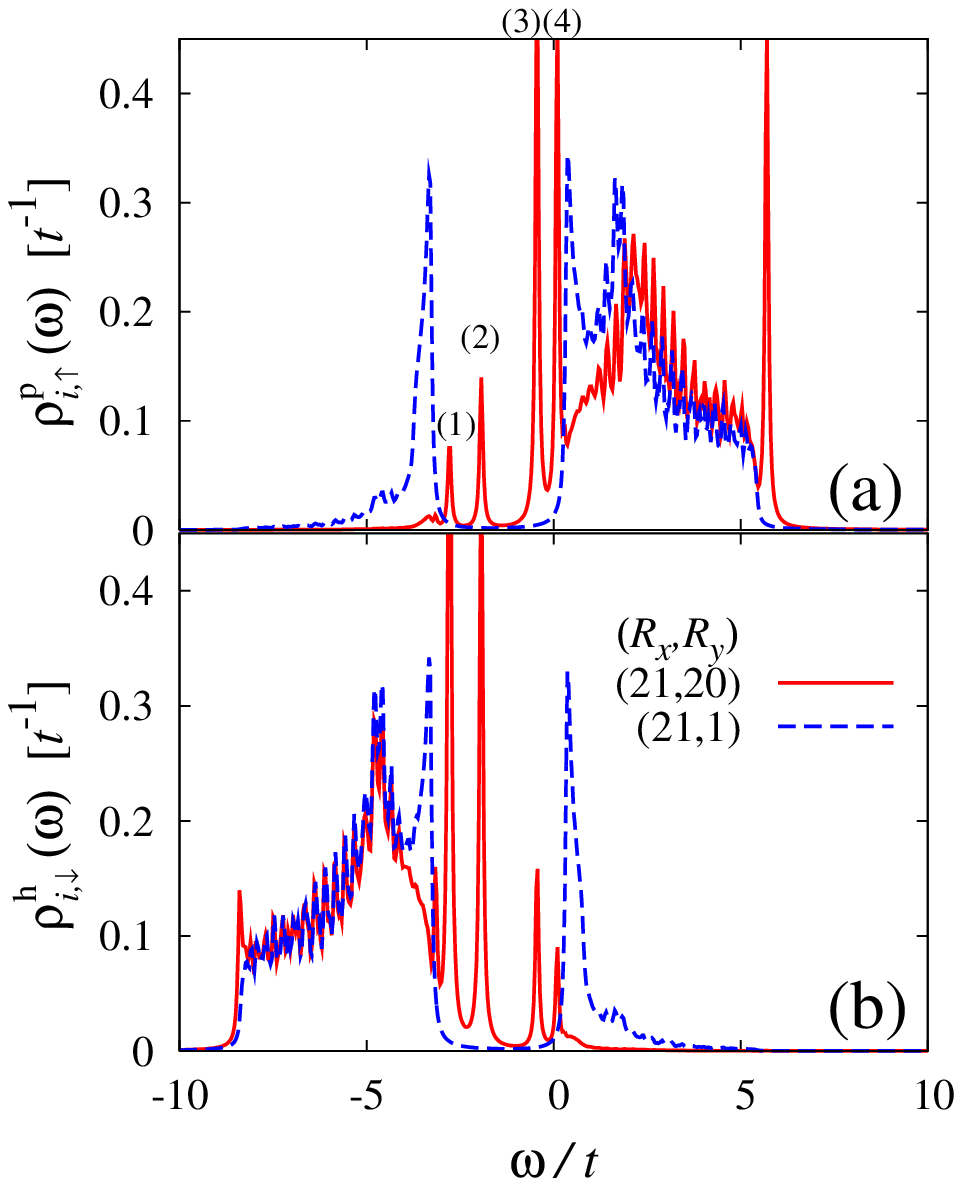}
\caption{(Color online) Local density of states in the polarized case ($N_\uparrow=301>N_\downarrow=300$). (a) $\rho^{\rm p}_{i,\uparrow}(\omega)$. (b) $\rho^{\rm h}_{i,\downarrow}(\omega)$. The BCS coherence peaks exist at $\omega=-3.32t$ and $0.38t$ far away from the impurity. (See the dashed lines in panels (a) and (b).)
\label{fig10}
}
\end{figure}
\par
\section{Localized excited states around magnetized impurity}

Figure \ref{fig8} shows the local density of states (LDOS) $\rho_{i,\uparrow}(\omega)$ around the impurity. In the unpolarized case (panel (a)), the impurity remains nonmagnetic, so that single-particle excitations are almost unaffected by impurity scatterings.  LDOS far away from the impurity (solid line in Fig.\ref{fig9}) agrees with the ordinary BCS density of states $\rho^{\rm p}_\uparrow(\omega)$ shown in Fig.\ref{fig2}. Even near the impurity, the overall structure of LDOS is unchanged (See the dashed line in Fig.\ref{fig9}.), except for a slightly smaller excitation gap, reflecting the suppression of $\Delta(i)$ around the impurity seen in Fig.\ref{fig5}(b).
\par
The situation is quite different in the presence of population imbalance. In this case, Fig.\ref{fig8}(b) shows that, in addition to the simple energy shift by the effective magnetic field $h=[\mu_\uparrow-\mu_\downarrow]/2=-1.47t$, LDOS is remarkably modified around the impurity. Since the bulk BCS coherence peaks are at $\omega=-3.32t$ and 0.38t (See the dashed line in Fig.\ref{fig10}(a).), the four peaks inside the BCS gap ($-3.32t\le\omega\le0.38t$) around the impurity seen in Fig.\ref{fig8}(b) are found to describe low-lying excited states induced by the magnetized impurity. Since these peaks disappear as one goes away from the impurity, they are bound states localized around the impurity. 
\par

\begin{figure}[t]
\includegraphics[width=14cm]{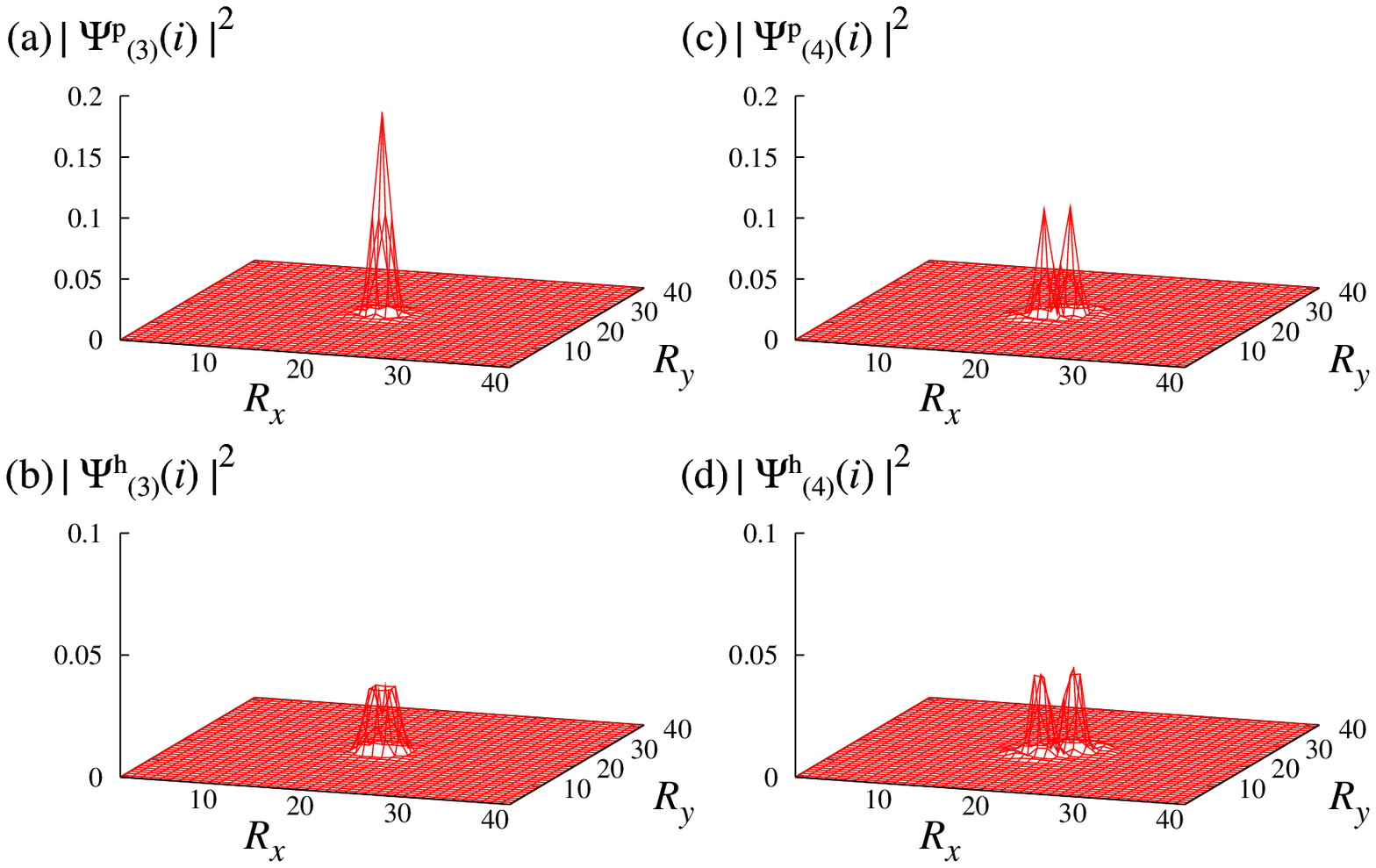}
\caption{(Color online) Calculated intensity of the wavefunctions of bound states. The upper and lower panels show the $\uparrow$-spin particle and $\downarrow$-spin hole components, respectively. Panels (a) and (b) show the bound state (3) in Fig.\ref{fig10}. Panels (c) and (d) show the bound state (4). We note that the right panels show one of the two degenerate bound states existing at the peak (4) in Fig.\ref{fig10}. For the other bound state, see Fig.\ref{fig12}(b2).
\label{fig11}
}
\end{figure}

These bound states also appear in the $\downarrow$-spin component of LDOS $\rho_{i,\downarrow}^{\rm h}(\omega)$, as shown in Fig.\ref{fig10}(b). The fact that all the peak positions inside the BCS excitation gap ($-3.32t\le\omega\le0.38t$) are the same between panels (a) and (b) means that they are composite excitations consisting of particle and hole components. To confirm this in a clear manner, as example, we show the wavefunctions $\Psi_j(i)=(\Psi_j^{\rm p}(i),\Psi_j^{\rm h}(i))$ of the bound states (3) and (4) in Fig.\ref{fig11}.
\par
In regard to the composite character of localized excited states, we note that the creation operator $\gamma_{{\bf p},\uparrow}^\dagger$ of the Bogoliubov excitation in the ordinary uniform BCS state is given by the sum of particle and hole creation operators as
\begin{equation}
\gamma_{{\bf p},\uparrow}^\dagger
=u_{\bf p}c_{{\bf p},\uparrow}^\dagger+v_{\bf p}c_{-{\bf p},\downarrow},
\label{eq.19}
\end{equation}
where $c_{{\bf p},\uparrow}^\dagger$ and $c_{-{\bf p},\downarrow}$ are the creation operators of $\uparrow$-spin particle and $\downarrow$-spin hole, respectively. That is, this basic property of ordinary single-particle Bogoliubov excitations is found to also hold in the bound states. 
\par
As discussed in Sec.II, $\rho^{\rm p}_{i,\uparrow}(\omega)$ and $\rho^{\rm h}_{i,\downarrow}(\omega)$, respectively, describe $\uparrow$-spin particle (hole) and $\downarrow$-spin hole (particle) excitations when $\omega>-h$ ($\omega<-h$). Using this, we find that the bound states (1) and (2) in Fig.\ref{fig10}, existing below $-h=-1.47t$, are composed of $\downarrow$-spin particle and $\uparrow$-spin hole excitations. Since the energies of the bound states (3) and (4) are larger than $-h=-1.47t$, they are found to be composite excitations of $\uparrow$-spin particle and $\downarrow$-spin hole.
\par
To see the mechanism of the bound state formation, it is helpful to examine the model Hamiltonian in Eq. (\ref{eq.2}) within the theory of magnetic impurity effect developed in the field of superconductivity\cite{Abrikosov,Maki,Shiba,Shiba2,Shiba3}, where the spatial variation of the superconducting order parameter is usually ignored. In addition to this simplification, when we also ignore the unimportant nonmagnetic potential ${\tilde V}(i)$, and approximate the magnetic impurity scattering to a $\delta$-functional potential, Eq. (\ref{eq.2}) reduces to, in momentum space,
\begin{eqnarray}
H=\sum_{{\bf p},\sigma}[\xi_{\bf p}-h\sigma]
c_{{\bf p},\sigma}^\dagger
c_{{\bf p},\sigma}
-\Delta
\sum_{\bf p}
\Bigl[
c_{{\bf p},\uparrow}^\dagger
c_{-{\bf p},\downarrow}^\dagger
+h.c.
\Bigr]
+{U \over 2}s_z\sum_{{\bf p},{\bf p}',\sigma}
\sigma c_{{\bf p},\sigma}^\dagger c_{{\bf p},\sigma}.
\label{eq.20}
\end{eqnarray}
Here, $s_z$ is the number of excess $\uparrow$-spin atoms localized around the impurity. When $s_z$ can take both positive and negative values depending on the direction of impurity spin, Eq. (\ref{eq.20}) is just the ordinary Hamiltonian describing a classical spin in a superconductor\cite{Abrikosov,Shiba}. Although the pseudospin in the present case always points to the $+z$-direction (when $N_\uparrow>N_\downarrow$), this difference is not crucial for the bound state problem. Indeed, Eq. (\ref{eq.20}) gives the same bound state solutions as those obtained in the ordinary magnetic impurity problem\cite{Shiba},
\begin{equation}
\omega=-h\pm
\Delta {1-[\pi\rho(0)s_zU/2]^2 \over 1+[\pi\rho(0)s_zU/2]^2},
\label{eq.21}
\end{equation}
where we have approximated the normal state density of states to the value $\rho(0)$ at the Fermi level, for simplicity. (For the derivation of Eq. (\ref{eq.21}), see the Appendix.) While four peaks exist inside the BCS gap in Fig.\ref{fig10}, Eq. (\ref{eq.21}) only gives two bound states. This means that the spatial variation of the order parameter ignored in Eq. (\ref{eq.20}) is crucial for the formation of bound state in the present case. 
\par

\begin{figure}[t]
\includegraphics[width=8cm]{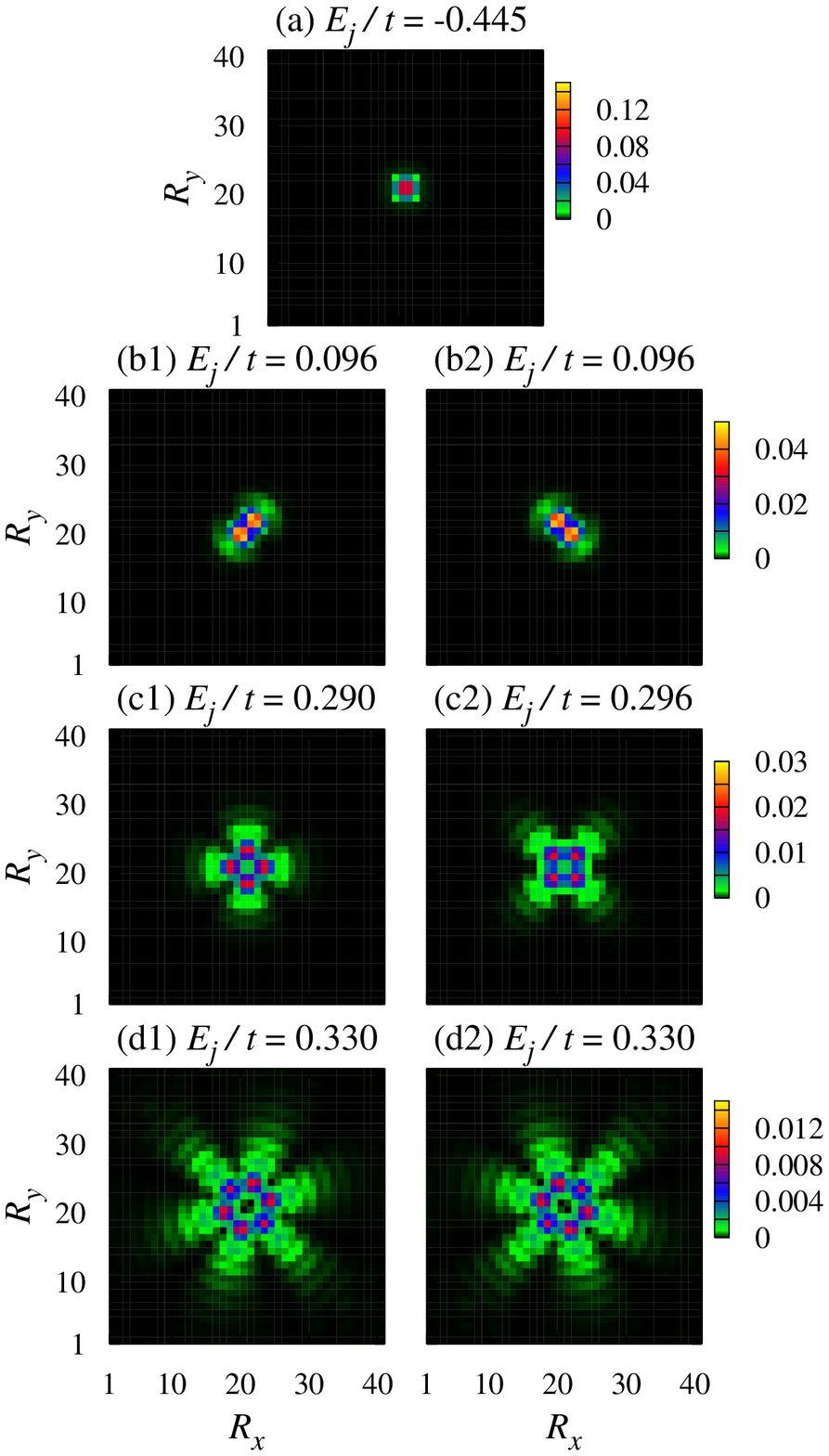}
\caption{(Color online) Intensity $|\Psi_j^{\rm p}(i)|^2$ of the bound state wavefunction with $\omega>-h$. Panel (a) shows the wavefunction of the bound state (3) in Fig.\ref{fig10}. The peak (4) in Fig.\ref{fig10} consists of doubly degenerate $p$-wave states given by panels (b1) and (b2). We also show other bound states with higher energies in panels (c) and (d) (although they do not appear as peaks in Fig.\ref{fig10}\cite{note3}). The reason why the degeneracy is lifted in panels (c1) and (c2) is the broken rotational symmetry by the background square lattice.
\label{fig12}
}
\end{figure}
\par
The superfluid order parameter $\Delta(i)$ is known to have a property similar to the ordinary potential\cite{Saint,Griffin}. Thus, the well structure of $\Delta(i)$ shown in Fig.(\ref{fig4})(d) makes us expect that the bound states seen in Fig.\ref{fig10} are trapped by this circular potential well\cite{note4}. Indeed, when we examine the bound states with $\omega>-h$, while the wavefunction of the lowest state (3) in Fig.\ref{fig10} is almost isotropic and non-degenerate (See Fig.\ref{fig12}(a).), the second lowest states (4) have the $p$-wave symmetry and are doubly degenerate, as shown in Figs.\ref{fig12}(b1) and (b2), which is consistent with the well-known result that the angular dependences of bound states in a circular potential well are given by $\cos(n\theta+\alpha)$ or $\sin(n\theta+\alpha)$ ($n=0,1,2,\cdot\cdot\cdot$). In Figs.\ref{fig12}(b1) and (b2), one finds $\alpha=\pi/4$, reflecting the breakdown of the rotational symmetry by the background square lattice. 
\par
We also obtain localized states with higher angular momenta, such as the $d$-wave ($n=2$) and $f$-wave ($n=3$) states, as shown in the lower four panels of Fig.\ref{fig12} (although they do not appears as peaks in Fig.\ref{fig10}\cite{note3}). Since their energies are close to the threshold energy $\omega=0.38t$ of the continuum spectrum, their wavefunctions spatially spread out, compared with the $s$-wave bound state in panel (a). Figure \ref{fig12} indicates that the magnetized impurity would be also useful for the study of a quantum dot in a superfluid Fermi gas. 

\begin{figure}[t]
\includegraphics[width=8cm]{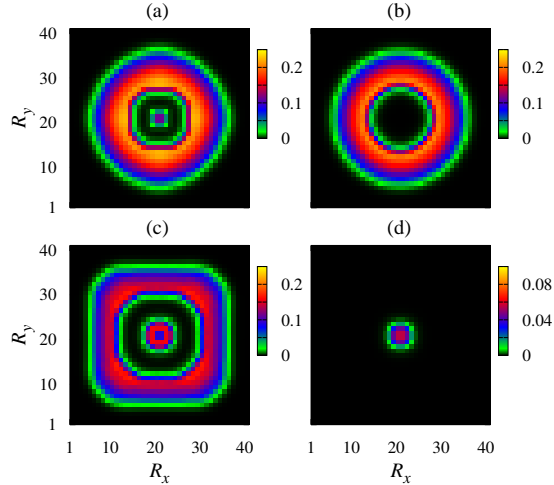}
\caption{(Color online) Intensity of local magnetization $s_z(i)$ in a trapped superfluid Fermi gas with population imbalance. Panels (a) and (b) show the case of harmonic trap $V_{\rm trap}^{(2)}(i)$ in Eq. (\ref{eq.22}). The population imbalance is chosen as (a) $(N_\uparrow,N_\downarrow)=(150,70)$ and (b) $(N_{\uparrow},N_\downarrow)=(150,80)$. Panel (c) shows the case of quartic trap $V_{\rm trap}^{(4)}$ in Eq. (\ref{eq.23}) with  $(N_\uparrow,N_\downarrow)=(150,80)$. Panel (d) shows the case of box-type trap $V_{\rm trap}^{\rm box}(i)$ in Eq. (\ref{eq.24}), where we set $(N_\uparrow,N_\downarrow)=(251,250)$ and $\lambda=2$. In all the cases, we take $V_{\rm trap}^0=5t$. For the impurity potential at the center of the system, we take $V_0/t=2$ and $\Gamma=1$.
\label{fig13}
}
\end{figure}
\par
\section{Effects of trap potential and finite temperature}
\par
So far, we have ignored effects of a trap, as well as finite temperatures, for simplicity. In this section, we briefly examine how they affect the localization of excess atoms.
\par
Figure \ref{fig13}(a) shows the intensity of local magnetization $s_z(i)$ in the case when the gas is trapped in the harmonic potential,
\begin{equation}
V^{(2)}_{\rm trap}(i=(R_x,R_y))=V_{\rm trap}^0
\Bigl[
\Bigl(
{R_x-R_{x0} \over R_{\rm max}-R_{x0}}
\Bigr)^2
+
\Bigl(
{R_y-R_{y0} \over R_{\rm max}-R_{y0}}
\Bigr)^2
\Bigr].
\label{eq.22}
\end{equation}
Here, $(R_{x0},R_{y0})=(21,21)$ is the center of the lattice, and $R_{\rm max}=41$ is the number of lattice sites in the $x$- and $y$-direction. In this panel, the finite intensity of $s_z(i)$ can be seen around the impurity at  $(R_{x},R_{y})=(21,21)$, which means that the magnetic impurity formation discussed in the previous sections also occurs in a trap. 
\par
However, when the population imbalance is lowered, all the excess atoms are localized around the edge of the trap, so that the impurity remains nonmagnetic, as shown in Fig.\ref{fig13}(b). Thus, the population imbalance must be large to some extent to obtain the magnetic impurity. 
\par
Since the excess atoms should avoid the spatial region where the potential is very large, the magnetization of the impurity is expected to occur more easily, when a trap potential around the edge of the gas is steeper than the harmonic trap in Eq. (\ref{eq.22}). Indeed, in the case of Fig.\ref{fig13}(b), when we replace the harmonic potential $V^{(2)}_{\rm trap}(i)$ by the quartic one\cite{Dalibard,Angom,Das},
\begin{equation}
V_{\rm trap}^{(4)}(i)=V_{\rm trap}^0
\Bigl[
\Bigl(
{R_x-R_{x0} \over R_{\rm max}-R_{x0}}
\Bigr)^4
+
\Bigl(
{R_y-R_{y0} \over R_{\rm max}-R_{y0}}
\Bigr)^4
\Bigr],
\label{eq.23}
\end{equation}
the localization of excess atoms is realized, as shown in Fig.\ref{fig13}(c). As a more extreme case, when we use the box-type trap\cite{Raizen},
\begin{equation}
V_{\rm trap}^{\rm box}(i)=V_{\rm trap}^0
\Bigl[
e^{-({R_x-1 \over \lambda})^2}
+
e^{-({R_x-R_{\rm max} \over \lambda})^2}
+
e^{-({R_y-1 \over \lambda})^2}
+
e^{-({R_y-R_{\rm max} \over \lambda})^2}
\Bigr],
\label{eq.24}
\end{equation}
excess atoms can be localized around the impurity even in the case of small population imbalance, as shown in Fig.\ref{fig13}(d). 

\begin{figure}[t]
\includegraphics[width=8cm]{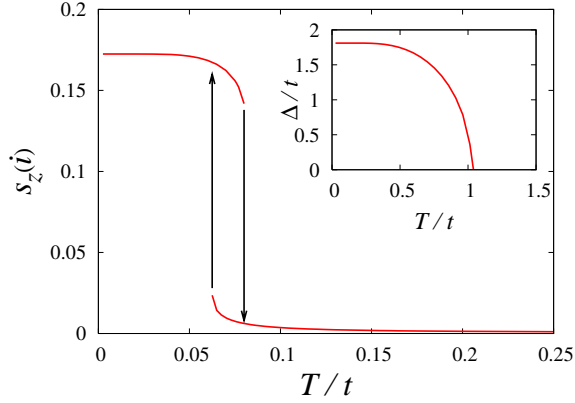}
\caption{(Color online) Local magnetization at the impurity site $s_z(i=(21,21))$, as a function of temperature. We take $N_\uparrow=301$ and $N_\downarrow=300$. The inset shows the calculated superfluid order parameter $\Delta$ as a function of temperature in the mean-field theory. In calculating $\Delta$ in the inset, we set $N_\uparrow=N_\downarrow=300$, and ignored impurity potential.  
\label{fig14}
}
\end{figure}
\par
Next we examine effects of finite temperatures within the mean-field theory\cite{note5}. Figure \ref{fig14} shows the local magnetization $s_z(i)$ at the impurity site, as a function of temperature. We find that the magnetization of the impurity remains finite even at finite temperatures. However, as shown in the inset, the present value of the pairing interaction $U=6t$ gives the mean-field superfluid phase transition temperature $T_{\rm c}/t\simeq 1$ in the absent of population imbalance. Thus, the sudden decrease of $s_z(i=(21,21))$ at $T/t\simeq 0.08$ means that the temperature must be far below $T_{\rm c}$ to realize magnetized impurities. 
\par
In Fig.\ref{fig14}, we see the hysteresis of magnetization, which is characteristic of the first order phase transition. Regarding the present magnetization phenomenon as a kind of phase separation, we find that this hysteresis behavior is consistent with the phase diagram of a polarized Fermi gas\cite{Shin,Ketterle3}, where the first order transition from the phase separating gas to the superfluid phase occurs far below $T_{\rm c}$ in the case of small population imbalance. 
\par
\section{Summary}
\par
To summarize, we have investigated how to realize magnetic impurities in a superfluid Fermi gas. In a two-dimensional attractive Fermi Hubbard model, we have calculated spatial variations of the superfluid order parameter, particle density, as well as polarization, around a nonmagnetic impurity within the framework of the mean-field theory at $T=0$.
\par
In the presence of population imbalance ($N_\uparrow>N_\downarrow$), we showed that the nonmagnetic impurity is magnetized in the sense that excess $\uparrow$-spin atoms are localized around it. This magnetized impurity behaves like an Ising-type magnetic scatterer, so that the superfluid order parameter is destroyed around it. This pair-breaking effect is similar to the magnetic impurity effect discussed in the superconductivity literature. 
\par
As another similarity between the present pseudo-magnetic impurity and real magnetic impurity in a metallic superconductor, the both impurities induce low-lying excited states below the BCS excitation gap. However, while the bound state formation by the latter magnetic impurity is usually discussed within the neglect of the spatial variation of the superconducting order parameter, we showed that the local suppression of the superfluid order parameter around the impurity is important in the former case. Namely, bound states are trapped inside a circular potential well formed by the superfluid order parameter. Because of this, the wavefunction of each bound state behaves like $\cos(n\theta+\alpha)$ or $\sin(n\theta+\alpha)$ ($n=0,1,2\cdot\cdot\cdot$) as a function of angle $\theta$. Thus, this magnetized impurity may be also viewed as a circular quantum dot in a superfluid Fermi gas. 
\par
Since the competition between magnetism and superconductivity is one of the most fundamental problems in condensed matter physics, our results would be useful for the study of this important topic by using superfluid Fermi gases. In this regard, we briefly note that the magnetic impurity discussed in this paper is a classical spin, because it does not have an exchange term (which is symbolically written as $s_+\sigma_-+s_+\sigma_-$, where $s_\pm=s_x\pm is_y$ and $\sigma_\pm=\sigma_x\pm i\sigma_y$ represent an impurity spin and the spin of a Fermi atom, respectively). As a result, the Kondo effect is absent. Thus, to examine the competition between the Kondo singlet and Cooper-pair singlet in cold Fermi gases, it is an interesting future problem to find out an idea to realize a quantum magnetic impurity with an exchange term in this system. 

\acknowledgments
This work was supported by Grant-in-Aid for Scientific research from MEXT, Japan (20500044, 22540412).

\appendix
\section{Derivation of the bound state energies in Eq. (\ref{eq.21})}
\par
To discuss the magnetic impurity problem in a Fermi superfluid, it is convenient to introduce the $4\times 4$ matrix thermal Green's function
\begin{equation}
{\hat G}({\bf p},{\bf p}',i\omega_n)=
-\int_0^\beta d\tau e^{i\omega\tau}\langle
T_\tau
\{
{\hat \Psi}_{\bf p}(\tau){\hat \Psi}_{{\bf p}'}^\dagger(0)
\}
\rangle,
\label{eq.a1}
\end{equation}
where ${\hat \Psi}_{\bf p}^\dagger=(c_{{\bf p},\uparrow}^\dagger,c_{{\bf p},\downarrow}^\dagger,c_{-{\bf p},\downarrow},c_{-{\bf p},\uparrow})$ is the four-component Nambu field\cite{Maki,Shiba}. In the absence of magnetic impurity, Eq.(\ref{eq.a1}) has the form ${\hat G}({\bf p},{\bf p}',i\omega_n)=\delta_{{\bf p},{\bf p}'}{\hat G}_0({\bf p},i\omega_n)$, where  
\begin{equation}
{\hat G}_0({\bf p},i\omega_n)=
{1 \over (i\omega_n+h)-\xi_{\bf p}\rho_3+\Delta\rho_1\tau_3}.
\label{eq.a2}
\end{equation}
Here, $\rho_i$ and $\tau_i$ ($i=1,2,3$) are Pauli matrices, acting on particle-hole space and spin space, respectively.
\par
The full Green's function in Eq. (\ref{eq.a1}) is obtained by including all magnetic impurity scatterings, which gives\cite{Shiba}
\begin{equation}
{\hat G}({\bf p},{\bf p}',i\omega_n)=
\delta_{{\bf p},{\bf p}'}{\hat G}_0({\bf p}.i\omega_n)
+
{\hat G}_0({\bf p},i\omega_n)
{\hat t}(i\omega_n)
{\hat G}_0({\bf p}',i\omega_n).
\label{eq.a3}
\end{equation}
Here, the $t$-matrix ${\hat t}(i\omega_n)$ involves effects of magnetic impurity smatterings as
\begin{eqnarray}
{\hat t}(i\omega_n)
&=&
{U \over 2}s_z\tau_3
+
\Bigl(
{U \over 2}s_z
\Bigr)^2
\tau_3{\hat F}(i\omega_n)+
\Bigl(
{U \over 2}s_z
\Bigr)^3
\tau_3{\hat F}(i\omega_n)^2+\cdot\cdot\cdot
\nonumber
\\
&=&
{\displaystyle {U \over 2}s_z\tau_3 \over \displaystyle 1-{U \over 2}s_z{\hat F(i\omega_n)}},
\label{eq.a4}
\end{eqnarray}
where ${\hat F}(i\omega_n)=\sum_{\bf k}{\hat G}_0({\bf k},i\omega_n)\tau_3$. We briefly note that, in the case of ordinary magnetic impurity problem where the impurity spin can take $s_z=\pm s$, we need to take the average over the direction of impurity spin, as $\langle s_z^{2n}\rangle=s^{2n}$, and $\langle s_z^{2n+1}\rangle=0$ ($n=0,1,2,\cdot\cdot\cdot$). In this case, ${\hat t}(i\omega_n)$ is given by\cite{Shiba}
\begin{eqnarray}
{\hat t}(i\omega_n)
=
{\displaystyle \Bigl({U \over 2}s\Bigr)^2 
\tau_3{\hat F}(i\omega_n)
\over 
\displaystyle 1-\Bigl({U \over 2}s\Bigr)^2
{\hat F}(i\omega_n)^2
}
.
\label{eq.a5}
\end{eqnarray}
\par
As usual, the factor ${\hat F}(i\omega_n)$ in Eq. (\ref{eq.a4}) can be conveniently evaluated by replacing the momentum summation by the energy integration. Assuming the particle-hole symmetry of the fermion band, and approximating the normal state density of states to the value $\rho(0)$ at the Fermi level, one obtains\cite{Maki,Shiba}
\begin{equation}
F(i\omega_n)=-\pi\rho(0)
{(i\omega_n+h)\tau_3+\Delta\rho_1 \over \sqrt{\Delta^2-(i\omega_n+h)^2}}.
\label{eq.a6}
\end{equation}
\par
Bound states are obtained as poles of the analytic continued $t$-matrix, ${\hat t}(i\omega_n\to\omega+i\delta)$. Using this, we obtain the equation for the bound state energies as
\begin{equation}
0={\rm det}
\Bigl[
1+{U \over 2}\pi s_z\rho(0)
{(\omega+h)\tau_3+\Delta\rho_1 \over \sqrt{\Delta^2-(\omega+h)^2}}
\Bigr],
\label{eq.a7}
\end{equation}
which gives Eq. (\ref{eq.21}). We briefly note that the bound state equation in the case of Eq. (\ref{eq.a5}) is given by
\begin{equation}
0={\rm det}
\Bigl[
1\pm {U \over 2}\pi s\rho(0)
{(\omega+h)\tau_3+\Delta\rho_1 \over \sqrt{\Delta^2-(\omega+h)^2}}
\Bigr],
\label{eq.a8}
\end{equation}
which also gives Eq. (\ref{eq.21}).



\begin{thebibliography}{99}
\bibitem{note} In Fig.\ref{fig6}(b), one sees the anisotropic spatial distribution of excess atoms, which is because of the broken rotational symmetry by the background square lattice.
\bibitem{note4} We note that, even in the unpolarized case, the local suppression of superfluid order parameter shown in Fig.\ref{fig5}(b) may induce bound states. In this case, however, the pair-potential well is shallow as $\Delta(1,1)=1.81t$ and $\Delta(21,21)=1.38t$, so that bound states only appear near the BCS gap edges. In the case of Fig.\ref{fig5}, the lowest bound state in the positive energy region appears at $\omega=1.68t$.
\bibitem{note3} The reason why these bound states cannot be seen in Fig.\ref{fig10} is that their energies are very close to the BCS gap edge at $\omega=0.38t$. In addition, since their peaks are broadened by the small imaginary part ($\delta=0.05t$) added to eigen energies in calculating LDOS, they cannot be distinguished from the continuum spectrum above the BCS gap in Fig.\ref{fig10}.
\bibitem{note5} At finite temperatures, we replace the step functions $\Theta(-E_j)$ and $\Theta(E_j)$ in Eqs.(\ref{eq.6})-(\ref{eq.8}) by the Fermi distribution functions $f(E_j)$ and $f(-E_j)$, respectively.
\end{thebibliography}
\end{document}